\documentclass[12pt]{article}
\usepackage{amsmath}
\usepackage{amssymb}
\usepackage{subfigure}
\usepackage{latexsym}
\usepackage{hyperref}
\usepackage[sort]{cite}

\renewcommand{\thefootnote}{\fnsymbol{footnote}}
\renewcommand{\thanks}[1]{\footnote{#1}} 
\newcommand{\starttext}{
\setcounter{footnote}{0}
\renewcommand{\thefootnote}{\arabic{footnote}}}
\renewcommand{\theequation}{\thesection.\arabic{equation}}
\newcommand{\be}{\begin{equation}}
\newcommand{\bea}{\begin{eqnarray}}
\newcommand{\eea}{\end{eqnarray}}
\newcommand{\beq}{\begin{equation}}
\newcommand{\ee}{\end{equation}}
\newcommand{\eeq}{\end{equation}}
\def\ba{\begin{eqnarray}}
\def\ea{\end{eqnarray}}
\def\bd{\begin{displaymath}}
\def\ed{\end{displaymath}}
\def\nn{\nonumber}

\begin{document}
\renewcommand{\theequation}{\thesection.\arabic{equation}}
\begin{titlepage}
\bigskip
\hskip 3.7in\vbox{\baselineskip12pt
\hbox{}
\hbox{hep-th/0608137}}

\bigskip\bigskip\bigskip\bigskip
\centerline{\Large \bf Studies Of The Over-Rotating BMPV Solution}
\bigskip

\bigskip\bigskip
\bigskip\bigskip
\bigskip
\centerline{\bf Lisa Dyson\footnote{dyson@berkeley.edu}}
\bigskip\bigskip
\centerline{\it Department of Physics University of California Berkeley, CA 94720 USA}
\bigskip
\centerline{\it Theoretical Physics Group, LBNL, Berkeley, CA 94720 USA}
\bigskip\bigskip\bigskip\bigskip\bigskip

\begin{abstract}
We study unphysical features of the BMPV black hole and how each can be resolved using the enhan\c con mechanism.  
We begin by reviewing how the enhan\c con mechanism resolves a class of repulson singularities which arise in the BMPV geometry when D--branes are wrapped on K3.  In the process, we show that the interior of an enhan\c con shell can be a time machine due to non-vanishing rotation.  
We link
the resolution of the time machine to the recently proposed resolution of the BMPV naked singularity / ``over-rotating" geometry through the expansion of strings in the presence of RR flux.
We extend the analysis to include a general class of BMPV black hole configurations, showing that any attempt to ``over-rotate" a causally sound BMPV black hole will be thwarted by the resolution mechanism.  We study how it may be possible to lower the entropy of a black hole due to the non-zero rotation.  This process is prevented from occurring through the creation of a family of resolving shells.   The second law of thermodynamics is thereby enforced in the rotating geometry - even when there is no risk of creating a naked singularity or closed time-like curves.  
\end{abstract}
\end{titlepage}
\starttext
\baselineskip=18pt
\setcounter{footnote}{0}

\newpage
\tableofcontents
\newpage
\baselineskip=18pt
\setcounter{footnote}{0}

\section{Introduction}
\subsection{Motivation}
The possibility of traveling back in time\footnote{Throughout this paper, when we
refer to time travel, we are referring to travel back in time.} was
introduced when Cornelius Lanczos and later W. J. van Stockum discovered a solution to Einstien's Equations which is generated by rotating pressureless matter \cite{lanczos,van}.   In this solution,
time-like loops that closed on themselves could be formed.  This showed
that within the context of Einstein's Equations, it is possible for a
trajectory to begin at one point in time and end at that same point.  Later, Kurt G\"odel studied similar solutions, but with non-zero cosmological constant - rotating dust in  Anti-de Sitter space \cite{godel}.  These solutions sparked a series of discussions within the scientific community which led to the discovery of other causality violating solutions to Einstein's Equations.  Numerous arguments were also put forth as to why these solutions are physically impossible.  See for example \cite{visser} and references therein for a review.

Recently, there has been increased interest in time travel in string
theory following the discovery by Gauntlett ${\it et, al}$ \cite{gghpr} of
supersymmetric realizations of the G\"odel universe.  In \cite{bghv}, Boyda
${\it et, al}$ applied Bousso's prescription of holographic screens \cite{bousso}
to the supersymmetric G\"odel universe. They found that for an inertial
observer, a causally safe region is carved out by holographic screens.
This led to a proposal that holography could play a key role in protecting
chronology. 
Numerous papers have continued the study of closed time-like
curves in string theory. 

Time-travel is not the only undesirable feature of solutions to General Relativity.  There are numerous solutions which have regions of infinite curvature.
The physical implications of singular solutions led Penrose to make a 'cosmic censorship' conjecture \cite{penrose1}.  According to his conjecture, if a singularity were created, it would not be causally connected to distant observers.  Instead, a black hole would be created.  Specifically, in the case of physically reasonable matter undergoing gravitational collapse, the singular regions that are created are contained in black holes.  This is the weak cosmic censorship conjecture.  The strong cosmic censorship conjecture asserts that time-like singularities never occur.  In that case, even an observer inside the horizon of a black hole never 'sees' the singularity \cite{penrose2}. 

As a quantum theory of gravity, we can ask what string theory has to say about geometries with naked singularities.  As it turns out, there are string theory solutions which naively appear to have singular regions unshielded by horizons.  However, upon closer inspection, it has been found that stringy physics comes into play to resolve these unphysical regions \cite{ks, ps, jpp}.  One of the resolutions \cite{jpp,jmpr} has been dubbed the 'enhan\c con' mechanism since along with the geometric resolution comes an enhanced gauge symmetry.  In that case, singularities appear in the naive geometry due to induced negative D(p-4) charge when Dp--branes are wrapped on K3 (p$\ge$4).  The singularity occurs when the volume of K3 shrinks to zero, but \cite{jpp,jmpr} showed that the geometry receives corrections when the volume takes on a stringy value.  In that case, the region where the naked singularity appears in the naive geometry can never by created.  The singularity is ``excised" (see \cite{singres} and reference therein for a review).

In \cite{me}, an attempt was made to combine the concept of singularity resolution with chronology protection.  It was proposed that the enhan\c con mechanism is a tool that
string theory also employs to correct causally unsound geometries.
The specific example of the BMPV black hole and the
corresponding dual geometries was discussed.  In the BMPV black hole, when the rotation
parameter exceeds a certain bound, a time machine is created.  The enhan\c con mechanism was applied to this geometry and it was argued that the time
machine is never created. Instead, the matter that would create the
time machine is restricted from traveling beyond a chronology protection
radius outside of the would-be chronology horizon.  In this proposal, since the matter that
is responsible for creating the closed time-like curves in the naive
geometry never reaches the region where they would exist, the time machine
can never be formed.  The corrected geometry is free of causal violations
and chronology is preserved.

Another unphysical feature of geometries can occur when a black hole is present.  Since black holes obey the laws of thermodynamics, an adiabatic process that would lower the entropy of a black hole would be unphysical.  Indeed, in \cite{jm, jl}, it was shown that the second law of thermodynamics could be violated for five dimensional black holes when the compact space is K3.  This can happen due to the negative D-brane charge that is induced as discussed above.  Specifically, the entropy of the five dimensional black hole made out of $N_1$ D1 branes,  $N_5$ D5--branes and $N_k$ units of momentum running along the effective string is $S = 2 \pi \sqrt{N_1 N_5 N_k}$.  When the D5--branes are wrapped on the K3, curvature couplings induces $-N_5$ units of D1--brane charge giving an entropy of  $S = 2 \pi \sqrt{(N_1 - N_5) N_5 N_k}$.  Thus, if we adiabatically drop D5--brane probes into the black hole with charge $\delta N_5$, the change in entropy is $\delta S^2 = 2 \pi (N_1 - 2 N_5) N_k \delta N_5$ which, for $N_5 > N_1 / 2$ can be negative.  Thus the second law of thermodynamics would be violated.  In \cite{jm, jl}, it was shown that probes which could naively lower the entropy of a black hole were prohibited from entering the black hole by the enhan\c con mechanism.

In the case of the BMPV black hole with $j$ units of angular momentum, a similar violation can occur.  The entropy is  $S = 2 \pi \sqrt{N_1 N_5 N_k - j^2 / 4}$.  Assuming that the rotation is carried by one of the charge types, $N_k$ \cite{ms}, we can consider adding charge ($\delta N_k$, $\delta j$).  The corresponding change in the entropy is $\delta S^2 = 2 \pi (N_1 N_5 \delta N_k - j \, \delta j / 2)$.   If the angular momentum satisfies the bound $\delta j \le 2 \delta N_k$ \cite{gh1}, the second law of thermodynamics may be violated when $j >  N_1 N_5$.  Notice that the bound $j \le N_1 N_5$ appears in a different phase of the geometry (i.e. not the black hole phase) \cite{mathur}, one in which angular momentum is carried by the D1 and D5--branes.  This implies that there is a region of parameter space, $2N_k \ge j \ge N_1 N_5$, where it may be possible to violate the second law of thermodynamics.  This can happen independent of the appearance of closed time-like curves (although closed time-like curves may be present).\footnote{Note also that $N_k >>  N_1 N_5 $ is the region of parameter space where the entropy of the black hole is derived in the CFT.}   

\subsection{Plan Of Paper}
In this work, we study the three above mentioned unphysical features of geometries - naked singularities, closed time-like curves and violations of the second law of thermodynamics.  We show how their resolution can be linked through the enhan\c con mechanism.  We begin by
reviewing singularity resolution via the enhan\c con mechanism in the context of the five dimensional black hole. 
In the process, we find that the interior of a class of enhan\c con geometries are time machines. 
In a dual picture, the time machine is made out of fundamental strings coupled to RR flux proportional to the angular momentum.  After applying the lessons that we
learned from the enhan\c con to this geometry, following the proposal of \cite{me} we argue that a shell emerges
outside of the chronology horizon beyond which the fundamental strings cannot travel.
In this case, the causality violating region can never be created and chronology is
preserved.   

Turning D--brane charge back on, 
we generalize the chronology protection result for the BMPV black hole \cite{me}, allowing for
configurations with varying charge.  We show that if we begin with a
causally sound BMPV black hole and attempt to add matter that would create
a time machine, a family of chronology protection shells appear outside of the
horizon beyond which the potentially causality violating matter cannot
travel.  Since the matter does not travel beyond this radius, the time
machine is never created and chronology is protected.\footnote{We note
here that our result naturally extends to the region behind the horizon
(when $R^2_{cp} < 0$ in our coordinates).  However, we will
restrict our discussion to the observable region outside of the black
hole.}

While studying the generalized BMPV geometry, we consider the charge configurations that can be
constructed which, when dropped into the black hole, can decrease its
entropy and thus violate the second law of thermodynamics.
We show that, not only does the chronology protection
mechanism save us from causality violations, but it also serves as an
enforcer of the second law of thermodynamics.  Probes that would lead to
violations of the second law are prohibited from passing through the
horizon by the chronology protection proposal.  We go on to show that even in the causally sound regime, it
may be possible to decrease the entropy of the black hole by dropping an
appropriately constructed probe into the horizon.  We apply the same
analysis used in the chronology protection discussion to show that these
probes are restricted from entering the black hole as well.  This provides a generalization to the results of \cite{jm,jl} where violations of the second law resulting from wrapping D-branes on K3 were resolved.

The outline of the paper is as follows:  In section 2, we review the BMPV
black hole. We discuss its five-dimensional form, D-brane configuration
and CFT dual. In section 3, we review the enhan\c con mechanism.  We focus
on the the supergravity analysis of the geometry as in \cite{jmpr}.  In
section 4 we discuss chronology protection.  First, we
consider the interior of the enhan\c con geometry in the limit of
vanishing D--brane charge in section \ref{sec_enh_ctc}. We study the time machine that results and show
how chronology is resolved (section \ref{sec_enh_cp}) through the expansion of fundamental strings in
the presence of RR flux (section \ref{sec_dual}). In section \ref{sec_bmpv}, we turn on D--brane charge to reproduce the
chronology protection result for the BMPV black hole studied in \cite{me}.  The naked singularity that results when $J^2/4 > Q_1 Q_5 Q_k$ is also resolved.
We then generalize the chronology protection result to include geometries
with causally sound BMPV black hole interiors in section \ref{sec_general}.
In section \ref{sec_second} we study the second law of thermodynamics.   We show how probes may be constructed which, when dropped
into the black hole, can lower its entropy, violating the second law of
thermodynamics.  We show how the resolution chronology protection mechanism kicks in
at just the right location to prohibit this from happening in section \ref{sec_second_ctc}. We go on to
show in section \ref{sec_second_wo}  that probes which would violate the second law of thermodynamics can
be constructed even when no causality violations would result.  These probes are also prohibited from entering the black hole,
generalizing the results of \cite{jm,jl}.

\section{The Geometry}\label{sec_geo}

\subsection{Five Dimensional Black Hole}

Let us begin by presenting the five dimensional rotating black
hole solution  \cite{bmpv, cvetic, myersbh, gh2, herdeiro}. We will
consider a black hole with three charges, call them $Q_1, Q_5$, and $Q_k$.
The metric for this geometry is
\begin{eqnarray}
ds^2  &=&  - (f_1 f_5 f_k)^{-{2 \over 3}}\Big[ dt + {J \over 2 r^2} \sigma_3 \Big]^2 + (f_1 f_5 f_k)^{1 \over 3} \Big[ dr^2 + r^2 d \Omega_3^2 \Big]
 \label{bhgeo}
\end{eqnarray}
where the $f_i$ are harmonic functions associated with the charges, $f_i =
1 + Q_i/r^2$.  In addition to the metric, this supergravity solution has
the following moduli and gauge fields under which the black hole is
charged,

\be
e^{-2\phi} = {f_5 \over f_1} \quad \quad \quad e^{2 \sigma} = { f_k \over f_5^{1/4} f^{3/4}} \quad \quad\quad e^{\sigma_i} = {f_1^{1/4} \over f_5^{1/4}}
\label{bhmoduli}
\ee
\be
A =  {1 \over f_1} \ (- {Q_1 \over r^2} dt + { J\over 2 r^2} \sigma_3\ )
\ee
\be
A^k =  {1\over f_k} \ (-{Q_k \over r^2} dt + { J\over 2 r^2} \sigma_3\ )
\ee
\be
B =  Q_5 \cos\theta d\phi \wedge d \psi +  {1 \over f_1} { J\over 2 r^2 } dt \wedge \sigma_3
\label{blackhole}
\ee
with $i=1,...4$.
 
In order to have a rotating black
hole that preserves supersymmetry, the causality-violating ergoregion that
is usually associated with Kerr-like solutions must be absent.  In order
for the ergoregion to be absent, the horizon of the black hole must be
static.  By analyzing the above metric in the near horizon limit, one
finds \be \omega_i =  {g_{\phi_i t} \over g_{\phi_i \phi_i }} \Big|_{r=0}
= 0 \ee So the horizon is not rotating.  The nonzero rotation parameter
does, however, have a non-trivial effect on the horizon.  Instead of the
standard spherical horizon geometry, the rotation introduces a squashing
parameter \cite{myersbh}.  The horizon is a squashed sphere.  Since the
squashing parameter depends on the angular momentum, $J$, one can show
that a sensible description of the horizon can break down if $J$ becomes
too large.

The BMPV geometry has closed time-like curves.  Closed time-like curves
exist for all values of $J$.  However, if $J$ is small enough, all closed
time-like curves are hidden behind the horizon.  When $J$ exceeds a
critical value, closed time-like curves appear outside of the horizon
leading to observable causality violations.  The over-rotating black hole
geometry was studied in detail in \cite{gh2}.

The causality constraint on $J$ can be seen explicitly by looking at the
angular components of the metric.  The chronology horizon, $R_{ch}$, is
the location where $g_{\phi_i \phi_i}$ vanishes.  Closed time-like curves
appear when $g_{\phi_i \phi_i}$ is negative.  This occurs if \be J^2
> 4 \,(r^2 + Q_1) (r^2 + Q_5) (r^2 + Q_k)\ . \ee Since the horizon is
located at the origin in these coordinates, an observer outside of the
horizon will see closed time-like curves if \be J^2 > 4\, Q_1Q_5Q_k \ .
\ee

The Bekenstein-Hawking entropy is given by \be S = { \pi^2 \over 2
G_5} \sqrt{ Q_1 Q_5 Q_k - { J^2 \over 4}} \label{entropy} \ee
This quantity can be imaginary if $J$ is too large.  This coincides with the presence of a time machine.  The horizon destabilizes when $J^2 > 4 Q_1 Q_5 Q_k$ (since the area is imaginary) and the singularity at $r=0$ is a naked singularity \cite{myersbh,gh2}. 

\subsection{Ten Dimensional D--Brane Geometry}\label{dbrane}

The ten-dimensional supergravity solution describes D1 and D5--branes with
momentum running along the effective D--string wrapped on an $S^1 $
\cite{herdeiro}.  The D5--branes are additionally wrapped on $T_4$ or K3.  In the case, of K3 naked
singularities of repulson type can appear.  
We discuss this in detail in section
3. The metric, RR field and the dilaton are

\begin{eqnarray}
ds^2 &=& {1 \over \sqrt{f_1 f_5}} \Big[ -dt^2 + {Q_k \over r^2} (dz - dt)^2 + dz^2 + { J \over r^2} \ \sigma_3 \ (dz - dt)\Big] \quad \quad  \quad \quad \quad \quad \nonumber\\
& & \quad \quad \quad \quad \quad  \quad \quad\quad \quad \quad \quad \quad + \ \sqrt{{f_1 \over f_5}}\, ds^2_{\cal{M}} +  \sqrt{f_1 f_5}\, \Big[ dr^2 + r^2 d \Omega_3^2\Big]
\label{fullgeo}
\end{eqnarray}
\be
C^{(2)} = {1 \over f_1} dt \wedge dz +  {1 \over f_1}{J \over 2 r^2 }\sigma_3 \wedge (dz - dt) + Q_5 \ \cos \theta \ d \phi \wedge d\psi
\label{rr}
\ee
\be
e^{-2 \Phi} = {f_5 \over f_1} \ .
\label{dsoln}
\ee
${\cal M}$ is $T^4$ or K3 and the charges $Q_1, Q_5$, $Q_k$ and $J$ are given by

\be Q_1 = { g \alpha'^3 \over V} N_1, \quad \quad   Q_5 = g \alpha' N_5,
\quad   \quad Q_k = { g^2 \alpha'^4 \over R_z^2 V} N_k, \quad \quad J = {
g^2 \alpha'^4 \over R_z V} j \label{charges} \ee where we have $N_1$
D1--branes, $N_5$ D5--branes, $N_k$ units of right moving momentum, and
the angular momentum is quantized in terms of integers $j$.  $V$ is the
asymptotic volume of $T^4$ or K3 and $R_z$ is the radius of the $S^1$.

This is a IIB supergravity solution with four supercharges.  Closed
time-like curves are an integral part of this geometry.  They can be
constructed by considering the curves with tangent vectors \cite{herdeiro}
\be l^{\mu}
\partial_{\mu} = \alpha \, \partial _{z} + \beta \, \partial_{\phi} \ . \label{cyclech} \ee
where we define $\sigma_3 = d \phi + cos\theta \, d\psi$ and $\phi$ is the Hopf fiber.  Since $z$ is a compact direction, as is required to preserve space-time
supersymmetry  and necessary to link this geometry to the black hole
\cite{tseytlin}, for certain values of $\alpha$ and $\beta$ curves of this
type can be closed.  A quick calculation of the proper length of these
curves show that they can be time-like in the region $r < R_{ch}$. The
causality violations are more explicit in the T-dual geometry \cite{me}
and as was shown in \cite{invariant}, closed time-like curves are invariant under
T-duality.

We can also consider the gauge theory that describes this configuration
\cite{entropy, bmpv}. If we take the size of the $S^1$ to be much larger
than the size of the $T^4$, the effective description is the 1+1
dimensional  field theory living on the world volume of the D1--branes.  One can show that the causality bound coincides with a unitarity bound in the dual field theory \cite{bmpv,herdeiro}

\section{Review Of Singularity Resolution}\label{sec_enh}

As discussed in the previous section, in
order to create a black hole we could wrap the D5--branes on a four torus
or on K3.  In this section, we will review some properties of D5--branes wrapped on K3.  Wrapping Dp--branes on K3
induces negative D(p-4)--brane charge.  This in turn leads to
naked singularities of repulson type \cite{repulson}.  In \cite{jpp,jmpr} it was found
that string theory resolves repulson singularities by using the enchan\c
con mechanism.  The enhan\c con of the D1 D5--brane system wrapped on K3
was studied in detail in \cite{jm,jmpr}.  Their results were generalized
to include non-zero rotation in \cite{jl}. Studying how the
enhan\c con resolves naked singularities will give us insight into how
string theory repairs geometries that violate causality \cite{me}. We will
study the results of \cite{jm,jl} in detail here.  Interestingly,
we will find that while the enhan\c con has the desired effect of
resolving naked singularities, it can create geometries with closed
time-like curves where none existed in the naive geometry.  We will use
this fact to explore how chronology can be resolved below.

Consider a geometry with $N_5$ D5--branes wrapped on K3 $\times S^1$ and
$N_1$ D1--branes wrapped on the $S^1$.  Wrapping the D5--branes
on K3 induces negative D1--brane charge equal to ${\tilde Q_1} = - { g \alpha'^3 \over V} N_5$. The supergravity
solution of this configuration is as in equation (\ref{fullgeo}), but with charges
$Q_5$ and $Q_1 \rightarrow Q_1 + {\tilde Q_1}$.  

Plugging these charges into the metric, it can be seen that a singularity appears when $f_1 = 0$.  This occurs at a radius \be r^2_r = { g \alpha'^3 \over V} (N_5 - N_1) \ . \ee  Since the horizon is located at the radius $r=0$ in these coordinates, the singularity is outside of the horizon for $N_5 > N_1$, and this geometry has the unphysical feature of having a naked singularity that is causally connected to observers outside of the black hole.  In order to see how the enhan\c con repairs the naked singularity, we can follow
the standard procedure of building the geometry by adiabatically bringing in the
objects that create the geometry from asymptotically far away.  A probe
calculation will show us if this is a consistent thing to do.
In \cite{jpp,jmpr} an equivalence was shown between the worldsheet and
supergravity analysis of probes that create the geometry.  The authors found that massless modes appear at a radius
larger than the repulson radius.  They argued that the naive geometry with
a repulson singularity is never formed. Instead, the objects that would
have created the singularity cannot travel beyond a special radius, dubbed
the enhan\c con radius, were new massless modes appear and geometry gets
corrected. The resulting true configuration is free of all naked
singularities.  We will review the resolution using the the supergravity
techniques discussed in \cite{jmpr,jm,jl}.

In an effort to construct the repulson singularity, we must determine if
it is possible to construct a geometry made up of $N_5$ D5--branes wrapped
on K3 $\times S^1$ and $N_1$ D1--branes wrapped on $S^1$ as above.  We
will see if this is possible by beginning with a geometry with $N_5 -
\delta N_5$ D5--branes and $N_1 - \delta N_1$ D1--branes and attempting to
bring in $\delta N_5$ D5--branes and $\delta N_1$ D1--branes from
asymptotically far away. As in \cite{jmpr,jm,jl}, we can study the
behavior of the this configuration in the supergravity picture by adiabatically collapsing 
a shell of D--brane charge from asymptotically far away and determining
if this is a consistent thing to do.  In order to do this, we patch
together two geometries at a radius $r_i $ which is the location of the
additional $\delta N_5$ D5--branes and $\delta N_1$ D1--branes and
consider what happens when we let $r_i \rightarrow 0$. The geometry for both
regions is of the same form as (\ref{fullgeo}), but we will need to make
the following substitutions: The exterior geometry has harmonic functions
$f_1 $ and $f_5 $ but with $Q_1 = g l^2 (N_1 - N_5) V_* / V $. For the
interior region, replace $f_1 $ and $f_5 $ with the harmonic functions
\bea
h_1 &=& 1 + {Q_1 - \tilde{Q_1} \over r_{i}^{2}} + {{\tilde Q_1 } \over r^2} \nn \\
h_5 &=& 1 + {Q_5 - \tilde{Q_5} \over r_{i}^{2}} + {{\tilde Q_5 } \over
r^2} \ . \label{harmonic_h} \eea Where \bd {\tilde Q_1 } = g\, l^2 \, {V_*  \over V} ( N_1 -
\delta N_1 - (N_5 - \delta N_5)) \ed \be {\tilde Q_5 } = g\, l^2\,  ( N_5
- \delta N_5 ) \ . \ee

The metric is smooth across the incision radius.  Any discontinuity that
appears in its derivative should be interpreted as a $\delta $ function
source of stress-energy. Applying the standard Israel geometry matching
techniques \cite{israel}, we find the stress tensor for the shell is $S_{\mu\nu}
= T_{\mu\nu} /\ 2 \kappa^2 \sqrt{G_{rr}}$
 with:
\begin{eqnarray}
T_{\mu\nu} & = & \left( \frac{f_1^\prime}{f_1} + \frac{f_5^\prime}{f_5} -
\frac{h_1^\prime}{h_1} -
\frac{h_5^\prime}{h_5}\right) G_{\mu\nu} , \nonumber \\
T_{\mu\phi_i} & = &  \left( \frac{f_1^\prime}{f_1} +
\frac{f_5^\prime}{f_5} - \frac{h_1^\prime}{h_1} -
\frac{h_5^\prime}{h_5}\right) G_{\mu\phi_i} , \nonumber \\
T_{ab} & = & \left(\frac{f_5^\prime}{f_5} -
 \frac{h_5^\prime}{h_5}\right) G_{ab} , \nonumber \\
T_{ij} & = & 0 \ . \label{stress-energy}
\end{eqnarray}
The indices $\mu$, $\nu$ denote the $t$ and $z$ directions; $a$, $b$
denote the $K3$ directions; $i$, $j$ denote the angular directions along
the junction three-sphere; and $2 \kappa^2 = 16 \pi G_N =(2\pi)^7 \ell_s^8
g_s^2$ sets the Newton constant.

From the stress-energy tensor, we find that the tension of the shell is
\be T_{shell} \propto {1 \over A_3} \left( \delta N_5 \tau_5 f_1  + \tau_1
(\delta N_1 - \delta N_5) f_5 \right)\ .  \ee

As in the prototypical D6--brane case, if we consider only D5--brane
probes ($\delta N_1 = 0$), the tension of the probe vanishes before the
repulson singularity is reached. This happens at a radius \be\label{r_e}
r_e^2 = g_s \ell_s^2 \frac{V_\star}{(V - V_{\star})} 2N_5 , \ee where $V_*
= (2 \pi l_S )^4 $ and $V$ is the asymptotic volume of K3. This radius is
precisely the location where the coordinate volume of K3, $V(r)
= V (f_1/f_5)$ as measured in Einstein frame takes on the stringy value,
$V_* $.  For $r
> r_e $, the tension of the D5--brane probe is positive.  At $r_e $, the D5--brane probe is
tensionless. Beyond $r_e $ the D5--brane probe would have negative
tension. Since this would be unphysical, the D5--brane probe cannot travel
beyond $r_e $. Since D5--branes cannot travel beyond $r_e$ and the enhan\c
con radius is greater than the repulson singularity radius, $r_e > r_r$,
the repulson geometry, which depends on D5--branes being able to travel
into the interior of $r_r$, cannot be created. Instead a shell of
D5--branes forms at $r_e $ and, in the limit of vanishig $Q_k$ and $J$, the interior geometry is just flat space.  In
the case of non-vanishing  $Q_k$ and
$J$, it was argued in \cite{jm,jl} that these charges decouple from the D5--branes at the enhan\c con radius and are free to travel to the origin.  Thus, the geometry has D5-branes sitting at $r_e$ and a non-trival geometry created by $Q_k$ and $J$ charge in the interior of the enhan\c con shell. The shell is also the location where massless modes
appear and the gauge symmetry is enhanced. Note also that the
stress-energy in the transverse directions, $T_{ij}$, vanishes, so there
are no transverse force acting on the shell and it is consistent to bring
it in from asymptotically far away.

For a shell made up of D1--branes, $\delta N_1 \ne 0$, $\delta N_5 = 0$, the tension is always
positive.  This implies that there is no obstruction to bringing in
arbitrary D1--brane charge.  This makes sense since the repulson
singularity is caused by the wrapping of the D5--branes on K3.  Likewise,
the tension of the shell is always positive when the number of D1--branes
equals or is greater than the number of D5--branes, $\delta N_1 \geq
\delta N_5 $. This would imply that we can bring in D5--brane charge only
as long as the D5--branes are appropriately ``dressed" with positive
D1--brane charge.  For all values of $\delta N_5 > \delta N_1$, a repulson
singularity exists in the naive geometry and an enhan\c con shell appears
in just the right location to repair the naive geometry.

\section{Resolving Time Machines}\label{sec_fund}
We have seen how string theory employs the enhan\c con mechanism to
resolve a class of singular geometries.  This was a pleasing
discovery because it provided an important example of how string
theory resolves physically unsound geometries as a fundamental
theory of quantum gravity should.  In an effort to determine if
string theory has a way of resolving causality violations, 
the enhan\c con construction was applied to an ``over-rotating" BMPV black
hole in \cite{me}.  It was argued that the above analysis has
an analogue in chronology violating geometries.  We
motivate this result further here by zooming in on the interior of
a type of enhan\c con geometry that was constructed in the previous
section.  We will discover that closed time-like curves exist for all values of the charges $Q_k$ and $J$.

\subsection{Causality Violations Inside The Enhan\c con}\label{sec_enh_ctc}
To begin, notice, as discussed in \cite{jm,jl}, the enhan\c con
only depends on D1 and D5--brane charges.  Since the momentum charges
$Q_k$ and $J$ do not play a role in creating the repulson singularity, it
is natural that they do not play a role in its resolution.  In \cite{jm,jl},
it was argued that the string momentum modes decouple from the D--branes
that make up the enhan\c con shell and are free to travel to the origin.
This is an interesting result because it allows for the existence of
closed time-like curves in the interior of the enhan\c con shell, even
when there were none in the original naive geometry.

Let us consider the limiting case where we only have D5--branes wrapped on
K3 and no additional D1--brane charge (e.g. the harmonic functions are constant in the interior, $h_i(r) = f_i(r_e)$ in equation (\ref{harmonic_h})).  We find that closed time-like curves exist for all
values of the charges $Q_k$ and $J$ in this limiting case.  We discover
that, even though the enhan\c con mechanism saves us from the embarrassing
situation of being able to construct a singular geometry unshielded by a
horizon, it plays a key role in creating an equally troubling geometry.

To see this explicitly, let us zoom in on the interior of the enhan\c con.  The metric is given by
\begin{eqnarray}
ds^2 &=& {1 \over \sqrt{f_1(r_e) f_5(r_e)}} \Big[ -dt^2 + {Q_k \over r^2} (dz - dt)^2 + dz^2 + { J \over r^2} \sigma_3 (dz - dt)\Big] \quad \quad\nonumber\\
& &  \quad\quad\quad\quad + \ \sqrt{{f_1(r_e) \over f_5(r_e)}}\, ds^2_{\cal{M}} +  \sqrt{f_1(r_e) f_5(r_e)}\, \Big[ dr^2 + r^2 d \Omega_3^2\Big]
\label{enhanconinterior}
\end{eqnarray}
This metric is supported by the RR potential
 \be C^{(2)} ={ J \over 2 f_1(r_e) r^2 }  \ \sigma_3
\wedge (dz - dt)  \label{rr2} \ee and a constant dilaton,
\be
e^{-2 \Phi} = {f_5(r_e) \over f_1(r_e)} \ .
\ee
For simplicity, we will re-scale this geometry to absorb the constants $f_i(r_e)$.  Since the re-scaled geometry is a IIB solution in its own right \cite{herdeiro}, we will leave out the added complication of the enhan\c con shell for the moment and consider the
solution for all values of $r$.  The re-scaled geometry is (using the same coordinate names although it is understood that the new coordinates have been re-scaled appropriately)
\be ds^2 =  -dt^2 + {Q_k \over r^2} (dz - dt)^2 + dz^2 + {J \over r^2}
\sigma_3 (dz - dt) + \  ds^2_{\cal{M}} +   dr^2 + r^2 d \Omega_3^2 \ . 
\label{momentum_geo}
 \ee 
 The RR potential is
 \be C^{(2)} ={ J \over 2 r^2 } \ \sigma_3
\wedge (dz - dt)  \label{rr2} \ee and the dilaton is constant.

Closed time-like curves are an integral
part of this geometry.  The chronology horizon is the positive real
solution to the equation \be 1 + {Q_k\over r^2} - { J^2 \over 4 r^6}
= 0 \ \label{momentum_ch}. \ee This equation has a non-zero and positive
solution for all values of $J$. It follows that $R_{ch}$ is always
positive and closed time-like curves exist for all radii $r< R_{ch}$.

What can we do about these chronology violations?  We can begin by
applying what we learned from the supergravity discussion in section
\ref{sec_enh} to this geometry following the proposal in \cite{me}.   Before we do, let us recall
what this geometry is made of by performing a T-duality along the $z$
direction. In the T-dual picture, the momentum
modes become winding modes. The resulting geometry is constructed out of
fundamental strings supported by RR flux proportional to
the angular momentum parameter $J$.  The full solution has the following
fields:

 \be ds^2  =  {1 \over f_s} \Bigg[-\Big( dt + {J \over
2 r^2} \sigma_3 \Big)^2 +
dz^2\Bigg]
  + ds^2_{{\cal M}} +
dr^2 + r^2 d \Omega^2_3
 \label{fundstring}\ee
 \bea B^{(2)} &=& {1 \over f_s} \left( dt + {J \over 2
r^2 }\sigma_3 \right)\wedge dz
\nonumber \
\\ C^{(1)} &=& {J \over 2 r^2 } \sigma_3 \nonumber \\ C^{(3)} &=& - {1 \over f_s}  {J \over 2 r^2 }
\sigma_3 \wedge dt \wedge dz \nonumber \\
e^{-2 \Phi} &=& f_s \label{fundstring_fields}\ ,\eea  where we have replace the harmonic function $f_k$ with its dual $f_s$ representing fundamental string charge $Q_s$.  Closed time-like curves
are invariant under T-duality \cite{invariant}, but one can easily perform a quick
computation on the angular directions of this geometry to confirm that the
causality horizon is at the same location as in equation
(\ref{momentum_ch}).  Thus we find that the fundamental string geometry has
causality violations for all values of $J$ in the region where $r <
R_{ch}$.

\subsection{Chronology Protection Sphere}\label{sec_enh_cp} We will now
return to the supergravity analysis of the geometry to study how the causality violations might be repaired.  Beginning with flat space, one can consider the thought experiment of creating the geometry by adiabatically bringing in charge from asymptotically far away as in section \ref{sec_enh}.  In the
limit of vanishing $J$, the BPS objects that make up the geometry are fundamental strings\footnote{We will study the IIA geometry in section \ref{sec_dual}, but will use continue to use the dual language here}.  With angular momentum turned on, the fundamental strings must be supported
by additional RR flux.  
As we bring in the strings coupled to the RR potentials from infinity, the discontinuity in the derivative of the metric that results represents the $\delta$ function source of stress-energy.  
We can determine the tension of the shell of charge by deriving the
stress-energy tensor as we did for the enhan\c con in equation
(\ref{stress-energy}). We do this by matching the geometry in (\ref{momentum_geo}) with flat space in the interior.  

In order to match
internal flat space with the string geometry, we must ``twist" our coordinates to ensure the metric is smooth across the boundary $r=R$.  
Defining $u=z-t$, the angular coordinate of the Hopf fiber in the interior ${\bar \phi}$ is twisted as follows: 
\be {\bar
\phi} = \phi + {J \over 2 R^4} \, u 
\label{transformation}
\ee 
This implies that $J$ must satisfy the quantization condition:
\be 
{J \over 2 R^4} = {N  \over R_z}
\label{quantization}
\ee
for some integer $N$ since $z$ is compact.  We must also re-scale $v = t+z$ as follows
\bea
{\bar v} &= &v + {1 \over 2 R^2} \left( Q_k - {J^2 \over 4 R^4} \right) u\nonumber \\
\label{coordtz}
\eea

Before calculating the stress energy tensor, let us write the metric in a more general form:
\be 
ds^2 = -dt^2 + dz^2 + K(r)\, (dz - dt)^2 + 2 H(r) \, \sigma_3 (dz - dt) +  dr^2 + r^2 d \Omega_3^2 + ds^2_{\cal{M}}
\label{general_interior} \ . 
\ee
where 
we have the following expressions for $K$ and $H$ outside of the shell, $r>R$, and inside the shell, $r<R$:
\bea
K_{out}(r) = {Q_k \over r^2}\ ;  & &   \ H_{out}(r) = {J \over 2 r^2}  \nonumber 
\eea
\be
K_{in}(r) = {1 \over R^2} \left( Q_k - {J^2 \over 4 R^4} \right) + {J^2 \over 4 R^8} \ r^2 \ ; \   \  \ \ H_{in}(r) = {J \over 2 R^4} \ r^2
\label{KandH}
\ee
\\
It is clear that the metric is continuous across the boundary $r=R$.  The stress tensor for the shell is
\bea  (2 \kappa^2)^{-1}S_{\mu\nu} dx^{\mu}dx^{\nu} &= &- \left( -K^{\prime}_{out} + K^{\prime}_{in} \right) (dz - dt)^2 + \left( H^{\prime}_{out} - H^{\prime}_{in} \right) (dz - dt) \ \sigma_3  \nonumber \\ 
&=& - {1 \over R} \left[ K + \left( {H \over R} \right)^2 \right] (dz - dt)^2 - 2 \left( {H \over R} \right) \, (dz - dt) \ \sigma_3 
 \ . \label{momentum-stress-energy}
\eea
where $\kappa^2 =  (8 \pi G_N)$.  Happily, the transverse components of the stress-energy tensor vanish, indicating that there are no transverse forces acting on the shell that would prevent us from constructing it.  From this, we can determine the energy density associated with the shell.  The momentum vector is given by
\be
P^{\mu} = \int \sqrt{-g} \ S^{\mu 0}\,  dV
\ee
The energy density associate with the time-like killing vector $\xi^{\mu} \partial_{\mu} = \partial_{0}$ is given by $\xi^{0} S^{\,  0}_{0}$.  From this, we find that the tension of our $\delta$ function source is 
\be T = T_0^{\, 0} = {1 \over
R^3} \left( Q_k - {J^2\over 4 R^4}\right) 
\label{stringtension}
\ee
where we have defined $T_{\mu\nu} = 2 \kappa^2 S_{\mu\nu}$.

Asymptotically, the energy density is of the form of a shell of dual
fundamental strings as expected, \be T \sim {Q_k \over Area(S^3)} \ . \ee
Locally, the energy density decreases by $J^2 /\ 4 R^4$. This effect is
crucial for repairing the causal sickness of the geometry and is due to the coupling to the RR flux.
The tension of the shell vanishes at a critical radius \be R_{cp}^4 = {J^2
\over 4 Q_k} \ .\label{Rcp} \ee  This is also the radius where our coordinate transformation simplifies: ${\bar z} = z$, ${\bar t = t}$.  For radii greater than $R_{cp}$, the tension of the
shell is positive.  For radii less than $R_{cp}$, the tension of the shell
is negative. 

How do we interpret this critical radius?  In the spirit of
the enhan\c con discussion in section \ref{sec_enh} and proposal of \cite{me} we argue 
that $R_{cp}$ is a critical radius beyond which our fundamental
strings coupled to RR flux cannot travel.  At this radius, non-trivial
physics comes into play correcting the naive geometry. If the fundamental
strings travel beyond $R_{cp}$, unphysical negative energy states would be
present. Since $R_{cp}>R_{ch}$, the matter that constructs our geometry is
never able to travel into the region where closed time-like curves would
be created. The causally sick region is never created and our usual notion
of chronology is preserved. The critical radius is the
chronology protection radius as proposed in \cite{me} in the
limit of vanishing D--brane charge.  

We can also rewrite the stress-energy tensor by using the quantization condition due to the geometric ``twist" (\ref{transformation}):
\be  T_{\mu\nu} dx^{\mu}dx^{\nu} =  - {1 \over R^3} \left( Q_k  +  {J N \over 2 R_z}  \right) (dz - dt)^2 - 2 \left( {N \over R_z} \right) \ R \ (dz - dt) \ \sigma_3  \label{} \ . \ee
The tension then is given by:
\be T =  {1 \over R^3}\left (Q_k - {J N\over 2 R_z} \right)
\  . 
\ee 
Vanishing tension implies 
\be
J = {2 Q_k \over N} \, R_z
\ee
for some $N$.  The angular momentum is maximal, $J_{max}/ R_z = 2 Q_k$, when $N=1$.  
In terms of the charge quantization given in (\ref{charges}), we have: 
\be T =  {4 G_5 / \pi R_z  \over R^3}\left (N_k - {N j \over 2} \right) \ . \ee 
From this, we find that $T = 0$ yields the following results:
\be N j = 2 N_k  \ , \quad \quad  j_{max} = 2 N_k\ . \label{maxj}\ee

\subsection{T-dual IIA Shell}\label{sec_dual}
In the previous section, we showed that the tension of the shell of charges that make up our time machine geometry vanishes at a critical radius.  We argued that this radius serves as a minimum value in moduli space beyond which the fundamental strings supported by RR flux cannot travel.  It is interesting to see what happens to our geometry in a T-dual configuration. 
If we perform a T-duality along the $z$ direction, the metric given in equation (\ref{fundstring}) can be written as:
 \be ds^2  =  {1 \over f_s(r)} \Bigg[-\Big( dt + H(r)\, \sigma_3 \Big)^2 +dz^2\Bigg]
  + ds^2_{\cal{M}} +
dr^2 + r^2 d \Omega^2_3
 \label{stringgeometry}\ee
with $f_s(r) = 1 + K(r)$ and with $K(r)$ and $H(r)$ as defined in (\ref{KandH}) for the inner and outer regions and with  T-dual charge $Q
_k \rightarrow Q_s$.   The T-dual geometry has F1--strings coupled to 1-form and 3-form RR potentials (proportional to $J$) for radii $r>R$, while the geometric twist of flat space in IIB gives rise to flux branes in IIA in the interior as was noted in \cite{gh1}.   We will derive the stress energy tensor in the IIA picture using language similar to \cite{gh1}.  

We can re-scale the geometry to absorb the $f_k(R)$ factor at the location of the shell.  This can be done quite simply with the following identifications: 
\bea {t} &=& f_s(R)^{{1\over 2}} \, {\tilde t} \nonumber \\
 { z} &=& f_s(R)^{{1\over 2}}  \, {\tilde z}\nonumber \\
 { J} &=& f_s(R)^{{1\over 2}}  \, {\tilde J}\nonumber \ .
\eea
If we define the functions
\be \gamma_{out} = {1 \over f_s(R)} {Q_s \over R^2}\ \ \quad \quad \gamma_{in} = {1 \over f_s(R)} {\beta^2 R^2 \over 4} \ ,\nonumber
\ee
we can rewrite $f_s$ in the outer and inner regions as:
\bea f_{s}^{out}(r) &=& 1 + \gamma_{out} \left({R^2 \over r^2} - 1\right)\nonumber \\ 
f_{s}^{in}(r) &=& 1 + \gamma_{in} \left({r^2 \over R^2} - 1\right)\nonumber
\eea
where we have defined $\beta = {\tilde J}/2 R^4$. 
With the metric written in this form, the stress-energy tensor is
\be  T_{\mu\nu} dx^{\mu}dx^{\nu} =  - {1 \over R} \left( \gamma_{out} + \gamma_{in}  \right) \left(dt + \beta R^2  \, \sigma_3\right)^2 + 4 \beta R \ \left(dt + \beta R^2 \, \sigma_3\right) \ \sigma_3  + {\left( \gamma_{out} + \gamma_{in}  \right) \over R} dz^2\label{TdualSE} \ . \nonumber 
\ee
A quick calculation of the tension of the shell gives
\be 
T= {1 \over R} \left( \gamma_{out} + \gamma_{in}  - 2 \beta^2 R^2  \right) =  {1 \over R}\left(\gamma_{out} - \gamma_{in} \right)
\ee 
Requiring that the tension of the shell is non-negative gives us the chronology protection condition, $T\ge 0$ or equivalently:
\be
\gamma_{out} \ge \gamma_{in}  \quad \Rightarrow \quad \ j \le 2 N_k \, .  
\label{gamma_cp}
\ee

\subsection{Resolving The BMPV Time Machine}\label{sec_bmpv}

Let us return to the BMPV black hole. From the enhan\c con discussion in
section \ref{sec_enh}, we saw that momentum modes decouple from D-brane charge when attempting to build the singular repulson geometry
\cite{jm,jl}.  We studied the geometry created by the momentum modes separately
above in the language of the dual fundamental strings coupled to non-trivial
RR-flux. We saw that closed time-like curves exist for all values of the
charges $Q_k$ and $J$. We further saw how chronology can be protected.  Following \cite{me}, we argued that the
fundamental strings expand in the presence of the RR flux to form a wall
at a chronology protection radius.  In that case, the strings cannot travel
beyond this radius and are never able to create the causality violating
geometry. Through the expansion of the objects that create the geometry to just the right radius, our usual notion of chronology can be preserved.

In the full BMPV geometry the condition for closed time-like curves to
exist is softened and in fact come along with another unphysical feature, a naked singularity \cite{gh2,myersbh}.  These unphysical features are present only when $J^2> 4 Q_1 Q_5 Q_k$. In
\cite{me}, an attempt was made to construct the BMPV geometry by bringing in the
charges that make up the geometry from asymptotically far away.  There was no obstruction to bringing in the D1 and D5--branes extended along the z direction, but, in the spirit of our discussion of the fundamental strings in the previous sections, it was proposed that the $Q_k$ and $J$ charges could not travel
beyond a chronology protection radius $R_{cp}$ which was the critical
radius where the tension of the shell of fundamental strings supported by
RR flux vanishes.  To get a supergravity description, we match the IIB geometry (\ref{fullgeo}) to an interior geometry with only D1 and D5--brane charge, 
\be
ds^2 = {1 \over \sqrt{f_1 f_5}} \Big[ -d{\tilde t^2} + d{\tilde z}^2 \Big]
+ \sqrt{{f_1 \over f_5}}\, ds^2_{\cal{M}} +  \sqrt{f_1 f_5}\, \Big[ dr^2 + r^2 d{\tilde \Omega}_3^2 \Big]
\label{d1d5geo}
\ee
where $d{\tilde \Omega_3^2}$ is the metric on a three sphere.  In order to match the coordinates in the interior to the exterior geometry the metric must be continuous across the boundary, the interior coordinates are related to the exterior coordinates via a similar geometric twist as in (\ref{transformation})
\be {\tilde
\phi} = \phi + {J \over 2 {\tilde R^4}} \, u
\label{transformation2}
\ee
with $u = z-t$ and ${\tilde R^4}= R^4 F_1 F_5$ is the scaled radius of the shell once D--brane charge is turned on.  We must also perform the coordinate transformation for ${\tilde v} = {\tilde t} + {\tilde z}$:
\bea
{\tilde v} &=& v + {1 \over 2 R^2} \Big( Q_k - { J^2 \over 4 {\tilde R^4}} \Big) u   \nonumber 
\eea 
As we saw when we just had $Q_k$ and $J$ charge turned on, in order to ensure that our internal geometry is just flat space with D1 and D5--branes, we require that 
\be
{J \over 2 {\tilde R^4}} = {N  \over R_z}
\label{scaledquantization}
\ee
for some integer $N$ since $z$ is compact.  Also notice that $t = {\tilde t}$ and $z = {\tilde z}$ when $Q_k = J^2/4 {\tilde R^4}$, similar to equation (\ref{coordtz}).  As can be anticipated, this happens precisely at the proposed chronology protection radius.

One can attempt to build the geometry with an adiabatically collapsing shell of fundamental strings coupled to RR-flux.  The shell has tension of the same form as in (\ref{stringtension}), but with the rescaled radius.  
\be T = T_0^0 = {2 \over
R^3} \left( Q_k - {J^2\over 4 {\tilde R}^4}\right) 
\label{fulltension}
\  . 
\ee
The tension of this shell is positive for $Q_k >  J^2/4 {\tilde R^4}$, zero when $Q_k = J^2/4 {\tilde R^4}$ and negative when $Q_k <  J^2/4 {\tilde R^4}$.  As discussed in the case of the enhan\c con and the time machine above, we conclude that the matter that makes up the shell cannot travel beyond the location where $Q_k = J^2/4 {\tilde R^4}$.  
If we attempt to adiabatically bring in the shell of charge beyond the critical radius
$R_{cp}$, the shell has negative tension.  
In the presence of D1 and D5--brane charge, the value of the radius of the proposed resolving sphere is given by 
\be 
R_{cp}^2 = {(Q_1 + Q_5) \over 2} \left[ -1 + \sqrt{1-{ 4\over Q_k (Q_1
+ Q_5)^{2}}\left(Q_1 Q_5 Q_k -  {J^2 \over 4}\right)}\right]\ .
\label{bmpv_rcp} \ee 
Since the chronology protection
radius is always greater than the chronology horizon, the resulting
geometry is free of all causal inconsistencies.  Also, the naked singularity that appears in the ``over-rotating" geometry due to the destabilization of the horizon (the naive area at $r=0$ is imaginary) is resolved since the interior geometry now only has D1 and D5--branes. 

We can rewrite the tension of the shell in terms of the charges using the quantization condition due to the geometric ``twist" (\ref{transformation2}):
\bea T &=&  {1 \over R^3}\left (Q_k - {J N\over 2 R_z} \right)\nonumber \\ 
&= & {4 G_5 / \pi R_z  \over R^3}\left (N_k - {N j \over 2} \right) \ . 
\eea  
Vanishing tension implies 
\be
j = {2 N_k \over N}  \ \quad \Rightarrow \quad \ \ j_{max} = 2 N_k\ . \label{maxj2}
\ee
for some integer $N$.  The bound on the angular momentum, $ j_{max} = 2 N_k$ agrees with the microscopic bound discussed in \cite{gh1} when the angular momentum is carried by one type of charge component.

To summarize, the argument is that when the angular momentum parameter exceeds the three charge bound but has not exceeded the single charge bound $2 N_k \ge j > 2 \sqrt{N_1 N_5 N_k}$ (which can occur when $N_k > N_1 N_5$), the supergravity description may actually give the correct asymptotic description of a domain wall configuration in which the angular momentum is carried by a single charge, $N_k$.  In that case, beyond a certain region in moduli space, the supergravity solution must be corrected. 
In fact, the supergravity solution is signaling that the geometry must be corrected by yielding unphysical behavior such as a naked singularity and closed time-like curves. 
In \cite{gh1} (see also \cite{drukker}) a similar domain wall argument was made.  In that case, the rotation is carried by two charges, one of which is $N_k$.  A solution was constructed which had an ``over-rotating" BMPV exterior and a G\"odel space interior.  The domain wall linking the two geometries was a supertube.  The resulting solution was free of causality violations and naked singularities.

\subsection{General BMPV Black Hole Interior}\label{sec_general}
We can ask what happens if we begin with a causally sound geometry and add
charge to the system that would create closed time-like curves.  We find that there is a natural generalization to the chronology protection mechanism which prohibits
causality violating probes from falling into the black hole. For each
probe, we argue that a wall emerges outside of the black hole at just the right
location to prohibit the flow of causality violating charge.  In this way,
chronology is protected for general BMPV charge configurations.

To show this explicitly, let us begin by considering the limiting case of
the causally safe BMPV black hole with angular momentum satisfying $4\,
Q_1 Q_5 Q_k - J^2 = 0$. We will add additional charge $\delta Q_k$ and
$\delta J$ to the system. Closed time-like curves will exist in the new
geometry if $(J + \delta J)^2
> 4 \  Q_1 Q_5 (Q_k + \delta Q_k)$ or, to first order, \be
\bigg(Q_1 Q_5 \delta Q_k - {J \delta J \over 2} \bigg) < 0\ .
\label{general_cond} \ee

Consider the following generalized metric \bea
ds^2 &=& {1 \over \sqrt{f_1 f_5}} \Big[ -dt^2 + dz^2 + K(r)(dz - dt)^2 + 2 H(r)\, \sigma_3 (dz - dt)\Big] \quad\quad\quad\nonumber\\
& &  \quad\quad\quad\quad\quad\quad\quad\quad+\ \sqrt{f_1 \over f_5}\, ds^2_{\cal{M}} +  \sqrt{f_1 f_5}\, \Big[ dr^2 +
r^2 d \Omega_3^2 \Big] \label{general_interior} \ . \eea Let us begin with
an interior geometry with \bd K(r) = {Q_k \over r^2} \ed \be H(r)  = {J
\over 2 r^2} \ee and consider bringing in additional charge, $\delta Q_k$ and $\delta
J$, adiabatically from asymptotically far away.  The metric for the exterior geometry is
represented by equation (\ref{general_interior}) with \bd K(r) = K_{ex}(r)
= { Q_k + \delta Q_k \over r^2} \ed \be H(r) =
H_{ex}(r) = {J + \delta J \over 2 r^2} \ . \ee

We will paste these geometries together at a shell of radius $R$. The
$\delta$ function source that results represents fundamental strings with
charges $\delta Q_k$ and $\delta J$ (supported by RR
flux). In order to ensure that the metric is continuous across the
boundary at a radius $R$, we must perform the following geometric ``twist": \be \phi' = \phi + {\delta J \over 2 R^4 F_1 F_5}
\, u \ee and (leading order) coordinate transformation
\be v' = v + {1 \over R^2} \left( \delta Q_k + {
J\delta J \over 2 R^4 F_1 F_5}\right)u \ee where $F_i = f_i(R)$ and the quantization condition becomes
\be
{\delta J \over 2 {\tilde R^4}} = {N  \over R_z}
\label{scaledquantization2}
\ee
with rescaled radius ${\bar R_4} = R^4 F_1 F_5$.
Once we have performed the coordinate transformation to match the two geometries,
the interior geometry can be expressed in the same form as equation (\ref{general_interior}), but with \bd K(r) = K_{in}(r) = {1\over
r^2}\left(Q_k  - {J \delta J \over 2 R^4 F_1 F_5}\right)  +  {1\over
R^2} \left( \delta Q_k + {J \delta J \over 2 R^4 F_1 F_5}\right) \ed \be H(r) = H_{in}(r) = {J\over 2 r^2} +
{\delta J \over 2 R^2}\left({r^2 f_1 f_5 \over R^2 F_1 F_5}\right) \ee The metric is now continuous across the surface $r= R$.

The discontinuity in the derivative of the metric will give us the
stress-energy of the $\delta$ function source of fundamental strings
coupled to RR flux. We find that the tension of the shell takes on the
following form \be T^{0}_{0} ={1 \over 2} \left( - K_{ex}' + {H_{ex}' H_{ex} \over f_1 f_5}
+ K_{in}' - {H_{in}' H_{in} \over f_1 f_5} \right) \ee To leading order, this
expression reduces to \be T = {1 \over R^3} \left(\delta Q_k - { J \delta
J \over 2 R^4 F_1 F_5} \left[ 1 + {1 \over 2 F_1}+ {1 \over 2
F_5}\right]\right) \ee where $R$ is the location of the shell.  The form
of the tension tells us that the shell contains local non-trivial charge,
$\delta Q_k$ and $\delta J$ as expected. Asymptotically, the tension is of
the form of a shell of fundamental strings, $T \sim Q_k /\ Area(S^3)$.
Locally, we get a correction to the energy density proportional to the
angular momentum parameter.
Also, the transverse components of the energy momentum tensor
vanish, $T_{ij} = 0$ where $\{i,j\} = \{\phi, \theta, \psi\}$, so
there are no forces acting on our shell in the transverse directions prohibiting its construction. 

Consider the following function: \bea
G(R)& = &R^8 + 2 (Q_1 + Q_5) R^6 + \left( (Q_1 + Q_5)^2 + 2 {1 \over \delta Q_k}\left[Q_1 Q_5 \delta Q_k - {J \delta J \over 2 }\right]\right) R^4 \nonumber \\
&&+  (Q_1 + Q_5) \left({Q_1 Q_5 \over 2} + {3 \over 2 \delta Q_k} \left[Q_1 Q_5 \delta Q_k - {J \delta J \over 2} \right]  \right)R^2 + {Q_1 Q_5 \over \delta Q_k} \left[Q_1 Q_5 \delta Q_k - {J \delta J \over 2} \right] \nonumber\\
\eea The tension is a positive multiple of $G(R)$. Studying the solutions
to $G(R) =0$ will tell us how the tension of the shell behaves. First,
define \be \alpha = Q_1 Q_5 \delta Q_k - { J \delta J \over 2 } \ee
\begin{itemize}
\item For $\alpha > 0 $, the tension of the shell is positive for all real radii $R$.  This can be seen clearly by the fact that $G(R)$ is always positive for real $R$.
\item For $\alpha = 0 $, the tension of the shell is positive when $R > 0$ and vanishes at the origin.
\item For $\alpha < 0 $, the tension of the shell is positive for radii greater than a critical value,
$\tilde R_{cp}$.  The tension vanishes at $R = \tilde R_{cp}$ and is
negative for $R< \tilde R_{cp}$.  This can be clearly seen by noticing
that $G(R) = 0$ has a unique real solution, $\tilde R_{cp}$. $G(R)$
increases monotonically when $R > \tilde R_{cp}$ and decreases
monotonically when $R < \tilde R_{cp}$.
\end{itemize}

How can we interpret this results?  If we wish to construct a time
machine we can begin with the limiting causally sound geometry with
charges satisfying $Q_1 Q_5 Q_k = J^2/4$ and add charges $\delta Q_k$
and $\delta J$.  If we attempt to bring in this shell of charges from
asymptotically far away, if $J \delta J < 2 Q_1 Q_5 \delta Q_k$, the shell
has positive tension for all radii outside of the horizon. If \be \alpha =
\bigg(Q_1 Q_5 \delta Q_k - {J \delta J \over 2} \bigg) < 0\
\label{cp_condition} \, ,\ee the tension of the shell vanishes at a
non-zero radius $ {\tilde R}_{cp}$. This is the same condition for the
appearance of closed time-like curves that we saw in equation
(\ref{general_cond}). When attempting to bring the shell in beyond ${\tilde
R}_{cp}$, its tension is negative. Following our discussion in the previous sections, we conclude that the fundamental strings coupled to RR
flux do not travel beyond ${\tilde R}_{cp}$ and chronology is
preserved.  Happily, this happens before closed time-like curves are ever
created.

For the maximally spinning BMPV black hole, we have discussed probes of the geometry which preserve causality and
those that would create closed time-like curves.  We
have argued that causally safe probes are free to travel within the BMPV
geometry by considering a shell of charge and thereby create the rotating black hole while
causality violating probes do not travel beyond a
chronology protection radius which is outside of the would-be chronology
horizon.  This generalizes the proposal of \cite{me} to account for
the creation of black holes or potential time machines with arbitrary
charges.  The claim is that we cannot ``over-rotate" a causally sound black hole.  The generalized proposal will be instrumental in section \ref{sec_second} where we will consider the entropy of a general BMPV black hole.
We will find that for the maximally rotating black hole, the chronology
protection condition corresponds to the condition of restricting charge
that would decrease the entropy of the black hole from entering the black hole and thereby enforces the second law of thermodynamics.  We will discover that the resolution mechanism will also serve as the enforcer of the second law of thermodynamics even
when there is no risk of creating closed time-like.

\section{Enforcer Of The Second Law Of Thermodynamics}\label{sec_second}
\subsection{With Chronology Protection}\label{sec_second_ctc}
Recall that the entropy of the black hole and D--brane system is \be
S = { \pi^2 \over 2 G_5} \sqrt{ Q_1 Q_5 Q_k - { J^2 \over 4}}
\label{entropy2} \ee
The entropy can be imaginary if $J$ is too large.  The condition for
imaginary entropy coincides with the presence of a time machine and a naked singularity \cite{gh2}.
However, precisely at the point when the entropy would be imaginary, $S^2$
is negative and therefore makes a positive contribution to $R_{cp}^2$ in equation (\ref{bmpv_rcp}) under the square root turning it on (i.e., making it real and positive).
So a time machine which would have imaginary entropy is not created.  Instead, we have a causally safe geometry where the region with closed time-like curves is removed.  This
also applies to the geometries constructed in section \ref{sec_general} with a generalized BMPV interior.

In fact, the connection between entropy and chronology protection is even
stronger.
With a negative contribution to the entropy coming from the parameter $J$,
it may be possible to drop a BPS probe into the black hole which would
decrease the horizon area and hence decrease the entropy of the system.
This would violate the second law of thermodynamics and our general
understanding of the physics of black holes.  We can study this process to
determine if there is some sort of an obstruction.

Consider dropping a probe with charges $\delta Q_k$ and $\delta J$ into
the black hole.  The corresponding change in entropy will be \be \delta S
= {4 \pi^2 \over 2 S} \left(Q_1 Q_5 \delta Q_k - {J \delta J \over 2}
\right) \label{cpentropy} \ee This quantity can obviously be negative.
Specifically, the entropy and horizon area of the black hole would
decrease if the charges satisfy the following equation: \be \bigg(Q_1 Q_5
\delta Q_k - {J \delta J \over 2} \bigg) < 0 \label{alpha2} \ . \ee
Interestingly, this is the same condition for causality violations to
appear in the limiting case of the maximally spinning black hole (\ref{general_cond}).  Happily, if this equation
is satisfied the chronology protection mechanism will kick in to prevent
the probe from reaching the horizon. Since the change in entropy is equal
to a positive multiple of $\alpha$ in equation (\ref{cpentropy}), if the change in
entropy is negative, we get a real and positive value for the chronology
protection radius, $\tilde R_{cp}$.  The resulting probe will not be able
to travel beyond that point, thus thwarting its ability to decrease the
entropy of the black hole.  So we see that the chronology protection
radius not only serves to prevent closed time-like curves from forming, but
also serves as an enforcer of the second law of thermodynamics.  This
result is a nice extension of the result in \cite{jm,jl} where the entropy
corrections due to the wrapping of D--branes on K3 was considered.

\subsection{Beyond Chronology Protection}\label{sec_second_wo} It is interesting to note that
what we have been calling the chronology protection radius $\tilde R_{cp}$ may
be non-zero even when closed time-like curves are not present.  To see
this explicitly, let us consider the full condition for causality
violations.  If we add charges $\delta Q_k$ and $\delta J$ to a general
causally sound black hole with charges $Q_1, Q_5, Q_k$ and $J$,  the
condition for closed time-like curves to form is \be Q_1 Q_5 ( Q_k+ \delta
Q_k) - {(J + \delta J)^2 \over 4}< 0 \ . \ee To leading order, this
reduces to \be \bigg(Q_1 Q_5 Q_k - {J^2 \over 4}\bigg) + \alpha < 0\ . \ee
If we assume that the initial configuration is causally sound, the first
term in parentheses is always positive.  Thus chronology violations can
only occur when the second term, $\alpha$, is negative as we have
discussed. It is when this occurs that our chronology protection mechanism
kicks in to prevent the appearance of closed time-like curves. However, it
is also possible to have a negative $\alpha$ while maintaining a sum that
is greater than zero.  This occurs when $Q_1 Q_5 Q_k - J^2 / 4 >\alpha $.  Although closed time-like curves are not present in
this case, recall from the entropy discussion that the change in entropy
can still be negative since $ \delta S^2\propto \alpha$ as seen in equation
(\ref{cpentropy}). Luckily, in this case, ${\tilde R_{cp}}$ is non-zero as we saw in
section \ref{sec_general}.  A shell is formed at radius ${\tilde R_{cp}}$ just in
time prevent probes that would lower the entropy of the black hole from
traveling beyond a radius outside of the horizon.  
Our
study of causality violations has led us to an interesting result for
BMPV black holes. We have found a mechanism that prohibits violations of
the second law of thermodynamics when those violation would occur due to non-vanishing rotation. 

\section{Conclusion}
In this paper, we have studied several unphysical features of the BMPV black hole and related geometries - naked singularities, closed time-like curves and violations to the second law of thermodynamics.  Beginning with a discussion of how the enhan\c con mechanism resolves a class of naked singularities \cite{jpp} due to Dp--branes wrapped on K3, we applied similar techniques to time machine and singular BMPV geometries to study the chronology protection proposal of \cite{me}.  We studied the time machine that can be created in the interior of a class of enhan\c con shells and argued that the resolution is a puffed-up shell of  fundamental string charge supported by RR-flux.  For the BMPV geometry, this provides an alternative to the ``over-rotating" black hole picture which would have a naked singularity and closed time-like curves.  The proposed resolution of \cite{me} is that the angular momentum is carried by a single charge when $2 N_k \le j < 2 \sqrt{N_1 N_5 N_k}$ and that the charge is smeared along a resolving sphere at just the right radius to remove the unphysical features of the geometry.  

We generalized the analysis to include resolutions with arbitrary BMPV interiors, showing that the resolution mechanism prevents us from over-rotating an otherwise causally sound BMPV black hole.  In the process, we found the interesting result that the same mechanism prevents violations to the second law of thermodynamics even when there are no naked singularities or closed time-like curves.  This generalizes the result in \cite{jm,jl} where violations to the second law of thermodynamics coincided with the appearance of naked singularities due to wrapping branes on K3.   

The enhan\c con was shown to be stable under supergravity perturbations in \cite{stability}.  We expect the stability of the resolving shell to hold in the context of chronology protection.  Also, it is interesting to note that in order to create the resolving shell, we performed a geometric twist which is T-dual to flux branes.  The non-local physics that results from twisting a geometry has been the subject of many papers in recent years (see e.g. \cite{Fluxbranes} for a discussion).  It would be interesting to study the present work in this context.  

\section*{Acknowledgments}
I would like to thank Tehani Finch, Ben Freivogel, Ori Ganor and Eric Gimon for useful comments and discussions.  This paper is dedicated to the loving memory of Arthur Edward Dyson.

\end{document}